\documentclass[journal,twoside,web]{ieeecolor}
\pdfoutput=1
\usepackage{generic}
\usepackage{cite}
\usepackage{amsmath,amssymb,amsfonts}
\usepackage{algorithmic}
\usepackage{graphicx}
\usepackage{algorithm,algorithmic}
\usepackage{hyperref}
\usepackage{times}
\usepackage{bm}
\hypersetup{hidelinks=true}
\usepackage{textcomp}
\newtheorem{theorem}{Theorem}

\newtheorem{definition}{Definition}
\newtheorem{lemma}{Lemma}

\newtheorem{assumption}{Assumption}
\newtheorem{remark}{Remark}
\newtheorem{example}{Example}
\renewcommand{\baselinestretch}{1.0}

\def\BibTeX{{\rm B\kern-.05em{\sc i\kern-.025em b}\kern-.08em
    T\kern-.1667em\lower.7ex\hbox{E}\kern-.125emX}}
\begin{document}
\title{Exponential Cluster Synchronization in Fast Switching Network Topologies: A Pinning Control Approach with Necessary and Sufficient Conditions }
\author{Ku Du, \IEEEmembership{Member, IEEE},
\thanks{Ku Du,
is with
the Artificial inteligience Department, Anhui University , Hefei,China
(e-mail: duku@ahu.edu.cn).
Yu Kang is with the Automation Department, University of Science and Technology of China,  (e-mail: kangduyu@ustc.edu.cn).
}}

\maketitle

\begin{abstract}
This research investigates the intricate domain of synchronization problem among
multiple agents operating within a dynamic fast switching network topology.
 We concentrate on cluster synchronization within coupled linear system under pinning control,
 providing both necessary and sufficient conditions. As a pivotal aspect,
 this paper aim to president the weakest possible conditions to
 make the coupled linear system realize cluster synchronization exponentially.
 Within the context of fast switching framework, we initially examine the necessary conditions,
 commencing with the transformation of the consensus problem into a stability problem,
 introducing a new variable to make the coupled system achieve cluster synchronization if the system is controllable;
 communication topology switching fast enough and the coupling strength should be sufficiently robust.
 Then, by using the Lyapunov theorem, we also present that the state matrix controllable is necessary
 for cluster synchronization. Furthermore, this paper culminating in the incorporation
 of contraction theory and an invariant manifold, demonstrating
 that the switching topology has an average is imperative for achieving cluster synchronization.
 Finally, we introduce three simulations to validate the efficacy of the proposed approach.
\end{abstract}

\textbf{\emph{Note to Practitioners}}-
\textbf{Intelligent vehicle agents consensus is an vibrant and interdisciplinary field which lies in devising effective strategies to ensure agreement or alignment among a network of autonomous agents.
However, due to each agent has limited information about the overall system. Communication constraints, such as bandwidth limitations or disturbances, can hinder the exchange of information, making it challenging for agents to synchronize their states. Hence, we delineate necessary and sufficient conditions that elucidate the interplay between the topology graph, coupling strength, and consensus mechanisms which is essential for advancing the capabilities of autonomous systems in dynamic and complex environments.}


\section{Introduction}\label{sec:introduction}
In recent years, the design and analysis of multi-agent systems
have gained significant attention due to their applicability in
various domains, such as robotics, communication networks,
 and autonomous systems\cite{olfati2007consensus}\cite{jadbabaie2003coordination}.
The achievement of multi-agent consensus represents a fundamental concept
within the domain of multi-agent systems
focusing on the cooperative coordination and alignment of multiple autonomous agents towards a common objective.
The term ``consensus" denotes the achievement of an agreement or convergence among
the individual agents in terms of a shared state or decision \cite{ye2018evolution}.
The goal of consensus is to enable these agents to reach
a unified decision or state, despite the inherent
decentralized and asynchronous nature of their interactions\cite{su2015distributed}.
This process is particularly relevant in various applications,
such as sensor networks, and social networks,
where collaboration among autonomous entities are  essential\cite{wang2015bounded}.
The synchronization of multiple agents
within such systems is a critical aspect,
influencing the overall performance and functionality.
In a multi-agent system, each agent operates
independently with local information and interacts with neighboring agents to exchange information\cite{qin2017robust}.

To date, large number of researches are focus on consensus.
In \cite{jiang2022multi}, the authors deal with the tracking problem
for linear coupled system. Both communication delays and varying control input
are considered in this paper.
By using the communication-delay-related observer,
the authors present the results that a closed-loop error system
will be proved stability via a Lyapunov–Krasovskii function
even the input delays are heterogeneous.
In \cite{Fontan2020}, interval consensus problem is introduced.
The authors show that if the intersection agents intervals is nonempty,
agents are required to converge to a shared value within the specified intersection.
In \cite{Zhai2016}, the resolution of consensus in a second-order system is addressed,
wherein it is assumed that the topology undergoes switching among a finite number of undirected graphs.
Then, the authors explore the distributed finite-time
consensus problem within second-order multi-agent systems (MAS),
considering the existence of bounded disturbances\cite{Yu2015}\cite{10375356}.
They introduce a sliding model aimed at attaining precise finite-time
consensus despite the presence of disturbances.
In \cite{nowzari2019event}, the authors provide an overview of event-triggered coordination for the
attainment of multi-agent average consensus. The authors underscore particular considerations
pertaining to assumptions regarding the capabilities of network agents and the
resulting characteristics of algorithmic execution.
This encompasses factors such as interconnection topology,
trigger evaluation, and the influence of imperfect information on the overall process.
In \cite{griparic2022consensus},
the authors consider the consensus problem which
the switching communication topology is governed by a trust-based algorithm.
The coupled system deal the communication with Albert-Barabasi probabilistic model.
This method provide both algebraic condition and communication quality to
deal with the connectivity control problem.
Then, the underwater vehicle formation control is developed under two independent stochastic
process. The authors design the controller following by the connectivity links\cite{jmse11010159}.

Some paper are considering the consensus problems under randomly switching topologies
\cite{You2013a,Wen2019a,Shang2016,9899415}.
In \cite{You2013a}, the authors studies the consensus problem with linear system
under Markov swithcing topology.
In \cite{Wen2019a}, the authors consider the heterogeneous nonlinear system,
output consensus can be achieved by an fractional-order observer.
In \cite{Kang2018}, cluster consensus is investigated for linear system under
directed interaction graph.
The authors investigate the output consensus problem with heterogeneous nonlinear systems,
which allow the communication delays presence in time-varying topology\cite{9899415}\cite{Lu2018}.
In \cite{Shang2016}, the authors deal with the stochastic consensus for linear system
with topology uncertainties. The authors also presented the upper bounded
of time delay.

There are many papers focus on the intricate dynamics of cluster
 synchronization in a multi-agent environment\cite{wu2008cluster}\cite{xia2011clustering}
 \cite{lou2014distributed}\cite{li2015multi}\cite{9905626},
 particularly under the influence of a rapidly switching network topology\cite{Sniffen1992A}.
In \cite{Gao2024}, the authors deal with cluster consensus problem
with discrete high-order systems where large number of adversaries
against achieving consensus.
They propose a high-order resilient cluster consensus algorithm, which
present necessary and sufficient conditions for achieving resilient cluster consensus.

In the rapidly evolving landscape of multi-agent systems,
the assurance of security has become a paramount concern\cite{zhang2021mrpt}.
The traditional models of trust, once deemed sufficient for ensuring the
integrity of networked interactions, are now challenged by the intricate
nature of contemporary cyber threats.
This shift in paradigm has necessitated a reevaluation of security architectures,
leading to the emergence of the zero trust Framework.
In \cite{ma2023intentional}, the authors address the synchronization of second-order multi-agent systems.
The coupled systems have intentionally time delay and will not achieve exponential synchronization
through a non-delay controller.
With the introduction of the proposed distributed protocol utilizing both present and previous position states,
the analysis identifies the crossing frequencies and crossing delays of the associated quasipolynomial for the delayed systems.
The zero trust framework fundamentally questions the conventional assumption of implicit trust
within networked environments. It asserts that trust should not be presumed,
but rather continuously validated, reflecting the dynamic
and context-dependent nature of trust in today's interconnected systems.
As a consequence, this paradigm shift has not only reshaped the principles
of cybersecurity but has also introduced novel challenges
and opportunities for the coordination and synchronization of multiple
agents operating within a networked ecosystem,
the authors deal with  Connected and autonomous vehicles
\cite{10217033}.

The historical development of research in this domain unveils a trajectory marked
by the application of diverse methodologies to address the
evolving challenges of securing multi-agent systems.
From early studies in cryptography to advancements in decentralized control theories,
the community has strived to adapt to the changing nature of threats and network dynamics\cite{malik2021proactive}.
Recent efforts have particularly focused on the application of
graph theory and system modeling to explore the intricacies of
networked interactions and security protocols\cite{cui2019extensible}.

Despite these advancements, the fast-paced evolution of modern network
topologies has introduced a layer of complexity that demands a fresh perspective.
The transition towards a zero trust framework has emphasized the need for robust
conditions that ensure cluster synchronization in the presence of rapidly
switching network configurations\cite{han2022distributed}. Consequently, questions pertaining to
the formulation of these conditions and the achievement of
consensus among agents in dynamic environments remain at
the forefront of current research challenges\cite{habib2019security}.

This paper endeavors to contribute to the ongoing discourse
by conducting a comprehensive examination of cluster synchronization
under fast-switching network topologies.
By tracing the historical trajectory of research
in multi-agent systems security and delineating the current gaps in knowledge,
we aim to provide a solid foundation for addressing the intricate
challenges posed by the intersection of dynamic network environments
and stringent security requirements.
The principal contributions of this paper can be summarized as follows:
\begin{itemize}
  \item Our investigation establishes the crucial role of controllability
  in linear systems to achieve
 cluster synchronization within the context of a fast switching network topology.
This represents the first demonstration of the
  essential necessity of controllability in the consensus dynamics of
  linear agents operating within a fast switching network topology.
  \item Utilizing a meticulously designed feedback matrix and a time-invariant quadratic Lyapunov function that incorporates a defined structural constraint, while capitalizing on the precompactness condition.
  We demonstrate the requisite of union graph has an average condition for the attainment of
      exponential consensus among linear agents. This extends the renowned finding outlined in reference
      \cite{jiang2022multi}\cite{Gao2024}
       and represents the initial elucidation of the imperative role played in ensuring cluster synchronization
        among coupled linear agents within the context  of a fast switching network topology\cite{9268958}.
  \item In the realm of addressing multiagent problems, we introduce a novel concept termed  average graph. This index possesses a distinctive attribute in that it elucidates the collaborative influence of system parameters and network topology in achieving consensus\cite{9268958}.
      Specifically, through the formulation of a well-designed feedback matrix and under the assumption of precompactness, we establish that global and uniform exponential consensus among linear agents is attainable provided that union graph has an average and controllability conditions are satisfied.
\end{itemize}

The rest of the paper is organized as follows: In section \ref{sec_preliminary}, we introduce
the zero trust environment, basic graph theory and contraction theory.
The system model present in this section. In section \ref{sec_MainResults},
 main results about cluster synchronization are presented. We give the simulation to  verify
our theories in section \ref{sec_simulation}. The conclusions are presented in \ref{sec_con}.

$Notation$: In this paper, the following notations and symbols are utilized to
represent various parameters, variables, and concepts.
 For matrices and vectors,  uppercase $R$
 and  lowercase  $r$   letters are used, respectively.
 The transpose of a matrix is indicated by $R^T$.
$\lambda_i(R)$ is the eigenvalue of matrix $R\in\mathbb{R}^{n\times n}$, which can be arranged
as $\lambda_{min}=\lambda_1\leq \lambda_2\cdots \leq \lambda_n=\lambda_{max}$.
$R>0$ stand for a positive matrix.
The symbol $I$ denotes the identity matrix with dimensions that are suitable for the given context.
The expression $diag\{s_1,s_2,\cdots, s_n\}$ represents a diagonal matrix,
where the $i-th$ diagonal element is denoted by $s_{i}$.
The Euclidean norm is denoted by $\| \cdot \|$.  $\otimes$ represent the Kronecker product.
$ker(R)$ denote the kernel space of a matrix $R$.
$span(V )$ denote the space spanned by set $V$.
$\vec{1}=[1,1,\cdots,1]^T$.

\section{Preliminary}\label{sec_preliminary}

\subsection{Zero trust environment}
The concept of a Full Trust Environment represents a traditional approach to network security.
This section explores the foundational principles of a Full Trust Model,
which relies on the assumption that entities within a
predefined network perimeter can be inherently trusted.
The Full Trust Environment operates on the belief that
once inside the network perimeter, entities are considered
secure without continuous authentication.
This model historically emphasizes a boundary-centric security strategy,
wherein internal users, devices, and systems are granted
broad access privileges based on their location within the
presumed secure perimeter.
In the realm of network security, the zero trust environment constitutes a
paradigm shift from traditional trust models.
This segment of the research delves into the foundational
aspects of the zero trust framework, emphasizing a fundamental
skepticism towards any entity, regardless of its origin or location
within the network\cite{ma2023intentional}. Zero trust assumes that both external and
internal elements pose potential threats, prompting a continuous
verification and authentication process for every device, user,
and transaction. This approach challenges the conventional notion
of establishing trust based on network location,
urging a more dynamic and adaptive security posture.
The discourse on the zero trust environment sets the stage
for understanding the intricacies of securing networked
systems in an era where threats are diverse, pervasive, and constantly evolving\cite{10217033}.
\begin{remark}
Here, the function $\gamma_{ij}(t)$ is defined as the trust metric for agents in
a complex communication framework.
It is shown that $\gamma_{ij}(t)$ is a function associated with agents
$i, j$  and the time  $t$.
\end{remark}

\begin{remark}
In a zero-trust environment, intelligent agents play a crucial role
in ensuring security and trustworthiness across diverse applications.
Intelligent agents,
equipped with advanced algorithms and decision-making capabilities,
are instrumental in implementing and enhancing security measures in this context\cite{zhang2021mrpt}.
\end{remark}

\subsection{Graph Theory}
The interactions among the linear systems are model as a undirected graph
The vertices represent various objects, such as people,
cities, or objects, while the edges represent the relationships between them.
The interactions among agents are modeled by a directed topology graph with order $N$ and  node set
$V = {1, 2, \ldots, N}$ and a directed graph of edges \cite{Shang2016}.
Let $G = (V, \varepsilon, A)$ denote a weighted digraph of $\mathbb{R}^{N \times N}$,
where $\alpha_{ij}$ represents the weight of the directed edge
$(j, i) \in \varepsilon \subseteq V \times V$,
and a weighted adjacency matrix
$A = [\alpha_{ij}] \in \mathbb{R}^{N \times N}$ satisfying
$\alpha_{ij} \neq 0$ if $(j, i) \in \varepsilon$
and $\alpha_{ij} = 0$ otherwise.
The Laplacian matrix denoted by $L = [l_{ij}]{N \times N}$,
associated with the graph $G$,
$l_{ij}$ defined through the following formulation.
$l_{ij}=-\alpha_{ij}$, and $l_{i j}=\sum_{k=1, k \neq i}^N \alpha_{i k}$.
Additionally, it is assumed that $\alpha_{ii} = 0$ for all $i \in V$ to prevent $i = j$.

In a directed graph, the directed path of graph $G$
denoted as $(h_1, h_2), (h_2, h_3), \ldots$.
For any two distinct agents $m, n \in V$,
If graph $G$ exists a directed path from agent $m$ to agent $n$,
We call the directed graph $G$ is strongly connected.
If a subset of the edges of the original graph that connects all the vertices without forming any cycles,
A digraph is considered to possess a directed spanning tree.

Partition the node set $V$ into $p$ clusters, that is $(V_1, V_2, \ldots, V_p)$,
where the subclusters satisfy $V_i \cap V_j = \emptyset$ for all $i \neq j$, where $i, j \in {1, \ldots, p}$ and
$\bigcup_{i=1}^{p} V_i = V$. Let $N_i = |V_i|$ represent the count of agents in $V_i$,
and $G_i$ denote the topology of cluster $V_i$.
Consider $\bar{j}$ as the subscript representing the subset to which agent $j$ belongs,
Please note that $\bar{j} = i$ if and only if $j \in V_i$.
Considering the zero-trust environment,
the adjacency matrix and Laplacian matrix can be mathematically articulated as follows:
 $\mathcal{A}_{z t}=\left\{\gamma_{ij}( t) \alpha_{i j}\right\}$,
$
\begin{aligned}
& \mathcal{D}_{z t}=\operatorname{diag}\left\{\sum_{j=1}^N \gamma_{1j}(t) a_{1, j}, \ldots \sum_{j=1}^N \gamma_{Nj}(t) a_{N j}\right\}
\end{aligned}
$
and
$$\mathcal{L}_{z t}=\mathcal{D}_{z t}-\mathcal{A}_{z t}=
\left[\begin{array}{ccc}
L_{11} & & 0 \\
\vdots & \ddots & \\
L_{q 1} & \cdots & L_{q q}
\end{array}\right].
$$

To simplify the symbols, we still use notation $\mathcal{A}_{z t}=[a_{ij}] $,
where $a_{ij}=\gamma_{ij}( t) \alpha_{i j}$

\subsection{Some lemmas}
Contraction theory serves as a mathematical framework employed
for the examination of stability and convergence properties within dynamical systems.
It furnishes a collection of tools and conceptual frameworks for scrutinizing
the dynamic evolution of systems over time.

\begin{lemma}\cite{Lohmiller2005}\label{lem1}
Consider a dynamical systems
$$
\dot{\mathbf{y}}=\mathbf{g}(\mathbf{y}, t)
$$
where $\mathbf{g}(y,t)$ is a trivial nonlinear function.
Let $\frac{\partial \mathbf{f}}{\partial \mathbf{y}}$ stand for the Jacobian matrix
of matrix $G$.
If $\Theta(\mathbf{y}, t)^{\top} \Theta(\mathbf{y}, t)$ is a positive definite metric matrix
and
\begin{equation}\label{eq_contrac}
\mathbf{G}=\left(\dot{\Theta}+\Theta \frac{\partial \mathbf{g}}{\partial \mathbf{y}}\right) \Theta^{-1}
\end{equation}
is a uniformly negative definite matrix,
then the trajectories of system converge exponentially to a same trajectory.
The convergence rate is $\left|\sup _{\mathbf{y}, t} \lambda_{\max }(\mathbf{G})\right|>0$.
Then, we call the system (\ref{eq_contrac}) is  contracting.
$\Theta(\mathbf{y}, t)^{\top} \Theta(\mathbf{y}, t)$ is the system contraction metric.
\end{lemma}

\begin{lemma}\cite{Lohmiller2005}\label{lem2}
For a smooth nonlinear system $\mathbf{g}(\mathbf{y}, t)$,
if $$
\mathbf{V} \frac{\partial \mathbf{g}}{\partial \mathbf{y}} \mathbf{V}^{\top}<\mathbf{0} \text { uniformly },
$$
then, all the trajectories synchronize to invariant  manifold $\mathcal{M}$.

More comprehensive case, for a constant invertible matrix $\Theta$ on $\mathcal{M}^{\perp}$ satisfy
$$
\boldsymbol{\Theta} \mathbf{V} \frac{\partial \mathbf{f}}{\partial \mathbf{y}} \mathbf{V}^{\top} \Theta^{-1}<\mathbf{0} \text { uniformly. }
$$
then, all the trajectories synchronize to  manifold $\mathcal{M}$.
\end{lemma}


\begin{remark}
This theoretical framework finds particular applicability
in the investigation of control systems, robotics, and optimization\cite{jadbabaie2003coordination}
\cite{ma2023intentional}.
At the core of contraction theory lies the fundamental concept of defining a contraction metric. This metric quantifies the manner in which distances between points in the state space of a dynamical system alter over time. In a system characterized by contraction, proximate trajectories exhibit convergence, indicating system stability. This notion of contraction draws an analogy to the geometric concept of contraction, wherein distances between points diminish\cite{Lohmiller2005}\cite{Prtial_contrac} .
\end{remark}

\begin{lemma}
Assume $A$ and $B$ are constant Hermitian matrices,
and let $ \lambda_i(A) $ and $ \lambda_i(B) $ denote their respective eigenvalues
arranged in non-decreasing order. The Weyl inequality asserts that for each index  $i$,
$$ \lambda_i(A + B) \leq \lambda_i(A) + \lambda_i(B).$$
\end{lemma}

\begin{remark}
This Weyl inequality has wide-ranging applications in various fields,
including quantum mechanics, optimization, and control theory,
providing valuable insights into the spectral properties of
Hermitian matrices and influencing the analysis of numerous mathematical
and scientific problems\cite{9268958}.
\end{remark}

\subsection{system model}
Contemplate a collective assembly of $N$ nodes subject to the dynamics of the ensuing generic linear system:
\begin{equation}\label{eq1}
  \dot{x}_i(t)=Ax_i(t)+Bu_i(t) \quad  i=1,\cdots,N.
\end{equation}
where $x_i\in \mathbb{R}^{n}$ is agent state; $A\in \mathbb{R}^{n\times n}$,
$B\in \mathbb{R}^{n\times n}$, $C \in \mathbb{R}^{n\times n}$ are constants
matrices; $u_i$ is controller for agent $i$.


Our objective is to guide the systems within same cluster toward to designated trajectory.
For this purpose, we select $p$ specific trajectories $(z_1(t), z_2(t), \ldots, z_p(t))$ that satisfy:

\begin{equation}\label{eq2}
  \dot{z}_i(t)=Az_i(t)  \qquad i=1,\cdots, p.
\end{equation}

\begin{definition}(Cluster Consensus)
Linear system (1) is saied achieve cluster synchronization if the
trajectory of (1) satisfying
\begin{equation*}
  \lim_{t\rightarrow \infty} \|x_i-s_i(t)\| =0 \quad \forall i \in V_{\bar{i}}
\end{equation*}
\end{definition}

We employ the subsequent algorithm to facilitate the convergence of linear systems towards achieving cluster synchronization:
\begin{equation}\label{eq_controller}
  u_i=K\left[\sum_{j=1}^N c_{i j} a_{i j}(t/\epsilon)\left(x_j-x_i\right)+c_{\bar{i}} d_i( t/\epsilon)\left(s_{\bar{i}}-x_i\right)\right]
\end{equation}
where $\epsilon$ is a positive constant, $c_{ij}$ is coupling strength, $i,j=1,2,\cdots,N$.
\begin{remark}
According to the controller (\ref{eq_controller}), one can find the time metrics are different.
The time metric of topology switching are bigger than the time metric of system dynamics
if $0<\omega <1$.
We can also find that the smaller $\omega$, the faster network topology switching.
\end{remark}

\begin{remark}
There are two time metric in controller (\ref{eq_controller}).
Our primary contribution lies in demonstrating that the coupled systems,
operating under a switching topology, can attain cluster synchronization.
Although the communication topology do not have a spanning tree at every fix time $t$.
Then, the authors will establish the proof that coupled agent with rapidly
changing topologies can synchronize to  cluster synchronization
under specific conditions, even the switching topology disconnected
for all instantaneously topologies.
\end{remark}

\begin{assumption}\label{ass_1}
The state matrix pair $(A, B)$ is said to be stabilizable.
\end{assumption}

\begin{remark}
There exist a positive definite matrix $P$ and a constant $\xi>0$ satisfy
$PA^T+AP-\xi PB^TBP=\xi P$.
\end{remark}

\begin{assumption}\label{ass_2}
The inter-cluster coupling strengths $a_{i j}$, $\bar{i} \neq \bar{j}$, satisfy the following equality:
$$
\sum_{j \in \mathcal{V}_{\ell}} a_{i j}=0, \quad \forall i=1, \ldots, N, i \in \mathcal{V} \backslash \mathcal{V}_{\ell}, \ell=1, \ldots, p.
$$
\end{assumption}
\begin{remark}
  The in-degree balanced condition, as extensively employed in existing literature [30], [33], serves as a descriptive framework characterizing the interconnection of distinct clusters.
 positively weighted couplings between nodes operate as a mechanism for synchronizing the nodes, while
 negatively weighted couplings between nodes belonging to distinct clusters function as a means of desynchronization.
\end{remark}

\begin{definition}
The union of dynamically switching topology $\mathcal{G}(t)$
between $t_0$ and $t_1$ is a graph whose node set and adjacency matrix
are $\mathcal{V}$ and $\overline{\mathcal{A}}=\left[\bar{a}_{i j}\right]$, respectively,
where $\bar{a}_{i j}=\int_{t_0}^{t_1} a_{i j}(\tau) \mathrm{d} \tau$.
The union topology graph can be induced from $\overline{\mathcal{A}}$.
\end{definition}

\begin{definition}
For a continuous, bounded Laplacian function $L(t)$, who has an average if the limit
\begin{equation*}
L^{\infty}=\lim_{t\rightarrow \infty} \frac{1}{t}\int_{t_0}^{t_0+t}L(\tau)d\tau
\end{equation*}
is well defined and there exists a decreasing bounded function
$\beta(t)$ such that
$$
\|\frac{1}{t}\int_{t_0}^{t_0+t} L(\tau)d\tau-L^{\infty} \|  \leq \kappa \beta(t)
$$
where $\kappa$ is positive constant.
\end{definition}



Then, we  have the error system as following:
\begin{equation}\label{eq_error}
  \dot{e}(t)=(I\otimes A)e(t)-(\tilde{L}\otimes BK)e(t)
\end{equation}
where $e(t)=x_i(t)-z_i(t)$; $\tilde{L}=\mathcal{L}_{zd} +D$, $D=diag\{c_1,c_2,\cdots,c_N\}$.

For the partition Laplacian matrix,
under the Assumption \ref{ass_2}-the in-degree balanced condition,
the row sum of $\mathcal{L}_{\ell k}, \ell \neq k$ which stand for
the inter-cluster couplings strength between cluster
$\mathcal{V}_i$ and cluster $\mathcal{V}_{j}$  is 0.
Then, the Laplacian matrix $\mathcal{L}$ can be transformed into following form:
$$
\mathcal{L}=\left[\begin{array}{cccc}
c_1 \mathcal{L}_{11} & \mathcal{L}_{12} & \cdots & \mathcal{L}_{1 p} \\
\mathcal{L}_{21} & c_2 \mathcal{L}_{22} & \cdots & \mathcal{L}_{2 p} \\
\vdots & \vdots & \ddots & \vdots \\
\mathcal{L}_{p 1} & \mathcal{L}_{p 2} & \cdots & c_p \mathcal{L}_{p p}
\end{array}\right] .
$$

According to the previous description,
$\mathcal{L}_{\ell \ell}$ is the Laplacian matrix deduced from digraph $\mathcal{G}_{\ell\ell}, \ell=$ $1, \ldots, p$.

\begin{assumption}\label{ass_3}
The average communication topology  $G_{\ell\ell}^{\infty}$ associate with its leader $z_i$ has a spanning tree.
\end{assumption}

Under Assumption 5, the existence of a positive diagonal matrix
$\mathbf{\Xi}$ is ensured, meeting the condition
$ \mathbf{\Xi} \mathbf{L}^\top + \mathbf{L}{\bar{\mathbf{\Xi}}}  > 0 $,
where  $\mathbf{\Xi} = \text{diag}(\sigma_1, \ldots, \sigma_p)$
for  $ \sigma_i > 0 $ and $ i = 1, \ldots, p $.

Remark 3: We adopt the same $\mathbf{\Xi}$  as delineated in \cite{qin2017robust},
hence, its explicit representation is omitted for brevity.

\begin{remark}
According to the Definition 2, we give the definiton of
set $\Sigma=\{ \mathcal{L}(t) | \lim_{t\rightarrow \infty} \frac{1}{t}\int_{t_0}^{t_0+t}L(\tau)d\tau \quad \text{has an average} \} $,
where $\int_{t_0}^{t_0+t}L(\tau)d\tau = [\int_{t_0}^{t_0+t}l_{ij}(\tau)d\tau ]$.
Obviously, the set $\Sigma$ is a compact set.
These results are important for the proof of Theorem \ref{theo_3}.
\end{remark}


\section{Cluster Synchronization With Linear System}\label{sec_MainResults}
This section is dedicated to providing rigorous proofs for the
necessary and sufficient conditions outlined for cluster synchronization.
We validate the theoretical foundations of our proposed cluster
synchronization approach. The proofs establish conditions
ensure the convergence of agents towards consensus despite the
challenges posed by the fast-switching network topology and the
stringent security constraints imposed by the zero trust framework.

\subsection{Necessary Conditions for cluster synchronization}
In this subsection, we delve into the mathematical requirements for achieving cluster consensus within the context of a linear system under the constraints of a fast-switching network topology. We explore the conditions that must be satisfied for agents within a cluster to synchronize their states despite the dynamic nature of the underlying network connections. By employing advanced control theory and leveraging the principles of linear system dynamics, we aim to provide insights into the essential prerequisites for successful cluster synchronization.


\begin{theorem}
Consider the linear dynamical system (\ref{eq1}) whose communication topology are zero trust network $G(t)$.
If Assumptions \ref{ass_1}  \ref{ass_2} and \ref{ass_3} is achieved globally, uniformly,
There exists a constant $\epsilon^{*}$, for all $0<\epsilon<\epsilon^{*}$,
coupling strength $c_i>\frac{\lambda_{min}(\Xi L_0^{\infty+L_0^{\infty}\Xi})}{\lambda_{min}(\Xi_iL_{ii}+L_{ii}^T\Xi_i)}$,
The interconnected system described by equation (1) will rapidly achieve cluster synchronization
in an exponential manner under of the controller(\ref{eq_controller}).
\end{theorem}

\begin{proof}
The error system of (\ref{eq_error})  can be transformed into following dynamics:
\begin{align*}
 \dot{e}=&[I\otimes A-L^{\infty}\otimes BK]e(t)\\
 &+((L^{\infty}-L(t/\epsilon))\otimes BK)e \\
   \triangleq &  \mathcal{A}e(t)-{M}e(t)
\end{align*}
where $\mathcal{A}=I\otimes A-L^{\infty}\otimes BK$,
${M}=(L^{\infty}-L(t/\epsilon))\otimes BK$.

Defined $\omega(t,\epsilon)=\int_{0}^{t} {M}e^{-\epsilon (t-\tau)} $
Then, matrix $I+\epsilon \omega(t,\epsilon)$, in invertiable,
and we have $\epsilon\|\mu(t,\epsilon)\|=\mathcal{O}(\gamma(\epsilon))$,
where $\gamma$ is a class-$\mathcal{K}$ function, see remark \ref{remark_3}.

Then, we introduce a new variable $e(t)=(I+\epsilon \omega(t,\epsilon))\omega(t)$.
For the sake of brevity in notation, we omit explicit mention of the arguments.
Denote $(I+\epsilon \omega)^{-1}=I+F$.
 The ensuing equality is then expressed as follows:
\begin{align*}
\dot{\omega}= &(I+\epsilon \omega(t,\epsilon))^{-1}e \\
   =&-(I+\epsilon \mu)^{-1}(M-\epsilon \mu) \omega(t)  \\
   & + (I+\epsilon \mu)^{-1}(A+M)(I+\epsilon \mu)\omega(t)  \\
   =&[-(I+F)(M-\epsilon \mu)+(I+F)(A+M)(I+\epsilon \mu)] \omega(t)  \\
   =&-(M+FM-\epsilon \mu-\epsilon F \mu) \omega(t)  \\
   & +(A+FA+M+FM)(I+\epsilon \mu)\omega(t)   \\
   =&(-M-FM+\epsilon \mu \epsilon F\mu +A+\epsilon A\mu  \\
    &+FA+\epsilon FA\mu+M+\epsilon M\mu +FM+\epsilon  FM) \omega(t)\\
   =&(I+F)\mathcal{A} \omega(t)+(I+F+\mathcal{A})\epsilon \mu +F\mathcal{A} \epsilon \mu +(\mu+FM)\epsilon \mu  \\
   =&(I+F)\mathcal{A} \omega(t)  +\gamma(t)(I+F)  \tilde{M}(t)  \omega(t)  \\
   \leq & \alpha \mathcal{A} \omega(t)+ \alpha \gamma(t)  \tilde{M}(t)  \omega(t)
\end{align*}
where  $\tilde{M}=\frac{1}{\gamma(t)}(I+\mathcal{A}+M)\epsilon \mu  $, $\alpha\triangleq\|I+F\|$.

We can establish that $ M(t, \epsilon)$ is uniformly bounded.
There exist a $\epsilon_1$, such that  $\alpha$ is also bounded for all $\epsilon<\epsilon_1$.
$\gamma(\epsilon) {M}(t,\epsilon) $ is a class K function.

Then, we consider the system
\begin{align}\label{sys_omega}
  \dot{\omega}(t) =& {\mathcal{A}} \omega (t)+\gamma(\epsilon) \tilde{{M}}(t,\epsilon) \omega (t)
\end{align}

We choose the candidate Lyapunov function
\begin{equation}\label{lyapunov}
V(t) =\omega^T(\Xi\otimes P)\omega.
\end{equation}
for system  $ \dot{\omega}(t) = \tilde{\mathcal{A}} \omega (t)$,
where $\Xi=diag\{ \Xi_1, \Xi_2,\cdots, \Xi_p\}$, $K=B^TP$.

According to the Wely's inequality, we have
$\lambda_i(\Xi L^{\infty}+L^{\infty}\Xi)>\lambda_i(\Xi diag\{c_1 L_{11}^{\infty},\cdots,c_p L_{pp}^{\infty}
+ diag\{c_1 L_{11}^{\infty},\cdots,c_p L_{pp}^{\infty} \Xi\})+\lambda_i(\Xi L_0^{\infty}+L_0^{\infty} \Xi)$,
where $L_0^{\infty}=L^{\infty}- diag\{c_1 L_{11}^{\infty},\cdots,c_p L_{pp}^{\infty}\}$.

As $c_k> \frac{\lambda_{min}(\Xi L^{\infty}_0+ L^{\infty}_0\Xi)}{\lambda_{min}(\Xi L^{\infty}_0+ L^{\infty}_0\Xi}$, we have
$\lambda_i(\Xi diag\{c_1 L_{11}^{\infty},\cdots,c_p L_{pp}^{\infty} + diag\{c_1 L_{11}^{\infty},\cdots,c_p L_{pp}^{\infty} \Xi\})+\lambda_i(\Xi L_0^{\infty}+L_0^{\infty} \Xi)>0$.
and $\lambda_i(\Xi L^{\infty}+L^{\infty}\Xi)>0$.
There exist a $\xi_1$ satisfy $\Xi L^{\infty}+L^{\infty}\Xi >\xi_1 \Xi$.

Then, we have
\begin{align*}
  \dot{V}&=\omega^T [\Xi\otimes (PA+A^TP)- (\Xi L^{\infty}+L^{\infty}\Xi)\otimes PBB^TP] \omega   \\
    &\leq \omega^T [\Xi\otimes (PA+A^TP)- \xi_1\Xi \otimes PBB^TP] \omega   \\
   &\leq -\xi_1 V(\omega)
\end{align*}

Then, we consider the Lyapunov function (\ref{lyapunov}) for system  (\ref{sys_omega}),
one can get
\begin{align*}
  \dot{V} =& 2\omega^T A(\Xi\otimes P)\omega -2\omega^T \gamma M (\Xi\otimes P)\omega  \\
   =&\omega^T(I\otimes A-L^{\infty}\otimes PBB^T)(\Xi\otimes P)\omega  \\
    & + \omega(\Xi\otimes P)(I\otimes A-L^{\infty}\otimes BK)\omega   \\
  =&\omega [\Xi\otimes (PA+A^TP)- (\Xi L^{\infty}+L^{\infty}\Xi)\otimes PBB^TP] \omega  \\
  & + \omega(\Xi\otimes P)(I\otimes A-L^{\infty}\otimes BK)\omega
\end{align*}


Obviously, the Lyapuhov function (\ref{lyapunov}) also satisfy following inequality:
\begin{align*}
&\left|\left| \frac{\partial V}{\partial \omega } \right|\right|  \leq \beta_1 \| \omega \|
\end{align*}
where $\beta_1$ is a positive constant.



Then, we choose the Lyapunov function $ V(t) =\omega^T(\Xi\otimes P)\omega$
for system $ \dot{\omega}=\alpha \mathcal{A} \omega(t)+ \alpha \gamma(t)  \tilde{M}(t)  \omega(t)$,
we have
\begin{align*}
  \dot{V}= & \frac{\partial V}{\partial t}+\frac{\partial V}{\partial \omega}\left(\alpha\tilde{\mathcal{A}} \omega (t)+\alpha\gamma(\epsilon) \tilde{{M}}(t,\epsilon) \omega (t)\right) \\
   \leq  & -\alpha\xi_1 \|\omega\|^2 +\alpha\|\frac{\partial V}{\partial \omega}\|  \|\gamma(\epsilon) \tilde{{M}}(t,\epsilon) \omega (t) \|  \\
   \leq& -\alpha\xi_1 \|\omega\|^2 +\alpha\beta_1  \|\gamma(\epsilon) \tilde{{M}}(t,\epsilon) \| \| \omega\|^2 \\
   \leq &  -\xi_2 V(\omega)
\end{align*}
where $\xi_2=\frac{\alpha\xi_1-\beta_1  \|\gamma(\epsilon) \tilde{{M}}(t,\epsilon) \|}{\lambda_{\max}(\Xi\otimes P)}   $.

As $\gamma(\epsilon)$ function is a class-K function, then there exist a $\epsilon_2$, such that
$\xi_2>0$, for all $0<\epsilon<\epsilon_2$.

Then, according to the comparison Principle \cite{HASSANK.KHALIL2002},
 system (1) can achieve exponential consensus under assumption 1-4, controller
(\ref{eq_controller}) and $0<\epsilon\leq \epsilon^{*}\triangleq \min\{\epsilon_1,\epsilon_2\}$.
\end{proof}

Then, we will prove the $\epsilon\|\mu(t,\epsilon)\|=\mathcal{O}(\gamma(\epsilon))$.
\begin{remark}\label{remark_3}
  proof of $\epsilon\|\mu(t,\epsilon)\|=\mathcal{O}(\gamma(\epsilon))$
\begin{align*}
   & \int_{0}^{t} Q(\tau)  e^{-(\epsilon(t-\tau))} d\tau  \\
  =&  \int_{0}^{t}  e^{-(\epsilon(t-\tau))} d(\int_{0}^{\tau}Q(s) ds)  \\
  =& e^{-\epsilon(t-\tau)}  \int_{0}^{\tau} Q(s)ds |_0^t \\
    &- \epsilon \int_{0}^{t} \int_{0}^{\tau} Q(s)ds e^{\epsilon (\tau-t)} d(\tau)  \\
  =& \int_{0}^{t} Q(s)ds  - \epsilon \int_{0}^{t} \int_{0}^{\tau} Q(s)ds e^{\epsilon (\tau-t)} d(\tau) \\
  =&\int_{0}^{t} Q(s)ds -\epsilon \int_{0}^{t} e^{\epsilon (\tau-t)} d(\tau)  \int_{0}^{\tau} Q(s)ds \\
  =&e^{-\epsilon t} \int_{0}^{t} Q(s)ds \\
  &-\epsilon \int_{0}^{t}e^{\epsilon (\tau-t)} (\int_{0}^{t}  Q(s)ds
  - \int_{0}^{\tau} Q(s)ds ) d\tau  \\
\end{align*}

One can also get
\begin{align*}
   & \epsilon \int_{0}^{t}e^{\epsilon (\tau-t)} (t-\tau)d\tau  \\
  = & \epsilon \int_{0}^{t}e^{\epsilon (\tau-t)} (t-\tau)d(\tau-t)  \\
  = & -\int_{0}^{t} (\tau-t)  d(e^{\epsilon (\tau-t)})   \\
  =& -(\tau-t)e^{\epsilon (\tau-t)}|_0^t+\int_{0}^{t} e^{\epsilon (\tau-t)}  d(\tau-t)  \\
  =&  -(\tau-t)e^{\epsilon (\tau-t)}|_0^t+ \frac{1}{\epsilon} e^{\epsilon (\tau-t)}|_0^t  \\
  =& te^{-\epsilon t} -\frac{1}{\epsilon}+\frac{1}{\epsilon}e^{-\epsilon t}
\end{align*}

We can get $\|\mu(t,\epsilon) \|\leq k t \sigma(t) e^{-\epsilon t}+ \epsilon \int_{0}^{t}e^{\epsilon (\tau-t)}k \sigma(t-\tau)(t-\tau)d\tau $. Then,
\begin{align*}
    &\epsilon\|\mu(t,\epsilon)  \|  \\
 \leq &\epsilon [kt\sigma(0)e^{-\epsilon t}+\epsilon k \sigma(0) \int_{0}^{t}e^{-(\epsilon(t-\tau))} (t-\tau) ]   \\
    =& \epsilon [kt\sigma(0)e^{-\epsilon t}+k \sigma(0) (te^{-\epsilon t }-\frac{1}{\epsilon} +\frac{1}{\epsilon} e^{-\epsilon(t-\tau)}     )   ]   \\
 \triangleq & \gamma(\epsilon)
\end{align*}
where $\gamma(\epsilon)$ is a bounded $\mathcal{K}$-class function.

\end{remark}

%
%
%
%

\subsection{Sufficient Conditions for cluster synchronization}
Building upon the necessary conditions outlined in the previous subsection, we extend our analysis to identify sufficient conditions for achieving cluster consensus. This involves a comprehensive examination of the interplay between the trust value function, the characteristics of the fast-switching network, and the inherent properties of the linear system governing agent dynamics. Our goal is to establish conditions under which not only the synchronization is possible but also to quantify the robustness of the proposed approach under varying network scenarios.

\begin{theorem}
Consider the linear dynamical system (\ref{eq1})
whose communication topology are zero trust network $G(t)$
under controller (\ref{eq_controller}).
Suppose that Assumptions  \ref{ass_2}, \ref{ass_3}
hold, Design $K = B^TP$. If the coupled system (1) achieved cluster consensus, then, coupled matrix $(A,B)$ is controllable.
\end{theorem}

\begin{proof}
We just assume the matrix pair $(A,B)$ not satisfies the stabilizability  condition.
Then, there exists a vector $v$, such that
$v^TA=\lambda_i v$, $v^TB=0$.

We have
\begin{align*}\label{eq_theo2}
  (I\otimes v)\dot{e}(t)= & (I\otimes v)(I\otimes A)e(t)-(I\otimes v)(\tilde{L}\otimes VK)e(t) \\
    =& (I\otimes v)(I\otimes A)e(t)\\
    =& (I\otimes \lambda_i v)e(t) \\
    =&\lambda_i (I\otimes v)e(t)
\end{align*}

Then, we have $v^Te_i(t)= v^T e^{\lambda_i t}e(0)$.
As the coupled system can achieve cluster synchronization exponentially,
we can get $v^Te_i(t)\rightarrow 0$, $t\rightarrow \infty$.
While if $e(0)$ is not equal to zero, $v^Te_i(t)$ will not achieve to zero.
which achieve a contraction with the cluster synchronization of coupled system.
\end{proof}


\begin{theorem}\label{theo_3}
  Consider the linear interconnected system (\ref{eq1})
communicating over zero trust network $G(t)$ under controller (\ref{eq_controller}).
Suppose that Assumptions \ref{ass_1}, \ref{ass_2} hold,
Design $c_i>\frac{1}{2\lambda_{min}(L^{*})}$, $i=1,\cdots, p$. If the
coupled system (1) achieved cluster consensus,
the average graph $G(t)$ has a spanning tree.
\end{theorem}

\begin{proof}
We design the flow-invarian concurrent synchronization subspace $\mathcal{M}$
which is associate with concurrently synchronized states.
In order to make all the system achieve to manifold $\mathcal{M}$,
Let define $V$ is a projection on $\mathcal{M}^{T}$,
where $\mathcal{M}^T$ is complementary space of subspace $\mathcal{M}$.

Assuming the average of  graph topology dose not have a spanning tree.
In fact, we can choose the manifold $\mathcal{M}$ as $ker(L^{*})$,
for Laplacian matrix $L^*$ such that $V^TL^{*}V>0$ and
$span(\vec{1})\subsetneqq span(V) $,
where $span(V)$ is the space spanned by set $V$.


We choose the projection matrix as $V\otimes I$.
According to Lemma \ref{lem1} and Lemma \ref{lem2},
one can get
\begin{align*}
& (V\otimes P ) [(I\otimes A)-(\tilde{L}\otimes BB^T)](V^T\otimes P)  \\
& +(V\otimes P ) [(I\otimes A)-(\tilde{L}\otimes BB^T)]^T(V^T\otimes P)  \\
  =& VV^T\otimes  (PA+A^TP)  - 2V \tilde{L}V^T\otimes PBB^TP    \\
  < & 0
\end{align*}
where  $c_i>\frac{1}{2\lambda_{min}(L^{*})}$.

As the set of $\Sigma$ is compact, there exist a series Laplacian matrices
$\{L(s)\|s=1,2,\cdots,  \}$  which have an average matrix $L^{*}$.
Obviously, $L^{*}\in \Sigma$.

Then, according to the Lemma \ref{lem1} and Lemma \ref{lem2},
one can get the coupled systems synchronize to the space of $\mathcal{M}$.
According to the assume the network topology is not connected,
then, $span(\vec{1})\subsetneqq  span(V)$.
Obviously, coupled linear systems (\ref{eq1}) will not achieve cluster synchronization.
\end{proof}

\begin{remark}
This paper prove the controllability of $(A,B)$ and topology
connectivity conditions are necessary for cluster synchronization of  coupled
linear system.
If the condition of $(A,B)$ is not controllable or the union
graph does't has an average, then there exists a non-trivial manifold
in which all the trajectories cannot leave this manifold,
thus, global cluster synchronization is impossible.
\end{remark}

\section{Simulation}\label{sec_simulation}
To validate the practical efficacy of our proposed cluster synchronization approach,
we conduct a series of simulations. This involves implementing the mathematical
model developed in Section III within a simulated environment that mimics
the dynamics of a real-world network. We analyze the performance of
the system under various scenarios, including different network topologies.
The simulation results serve to corroborate the theoretical
findings and provide valuable insights into the
robustness and adaptability of the proposed synchronization strategy.

\begin{example}
We consider a collection of N agents connected through an over switching network topology.
The network topology is represented by a directed graph,
where the agents are the nodes, and the edges represent the
 communication links between the agents.
 We consider the multiagent systems consist of 7 agents,
 that is, $N=7$.
The communication topologies of  coupled
system (\ref{eq1}) are shown in Fig.\ref{fig1}.

\begin{figure}[!t]
\centerline{\includegraphics[width=9cm]{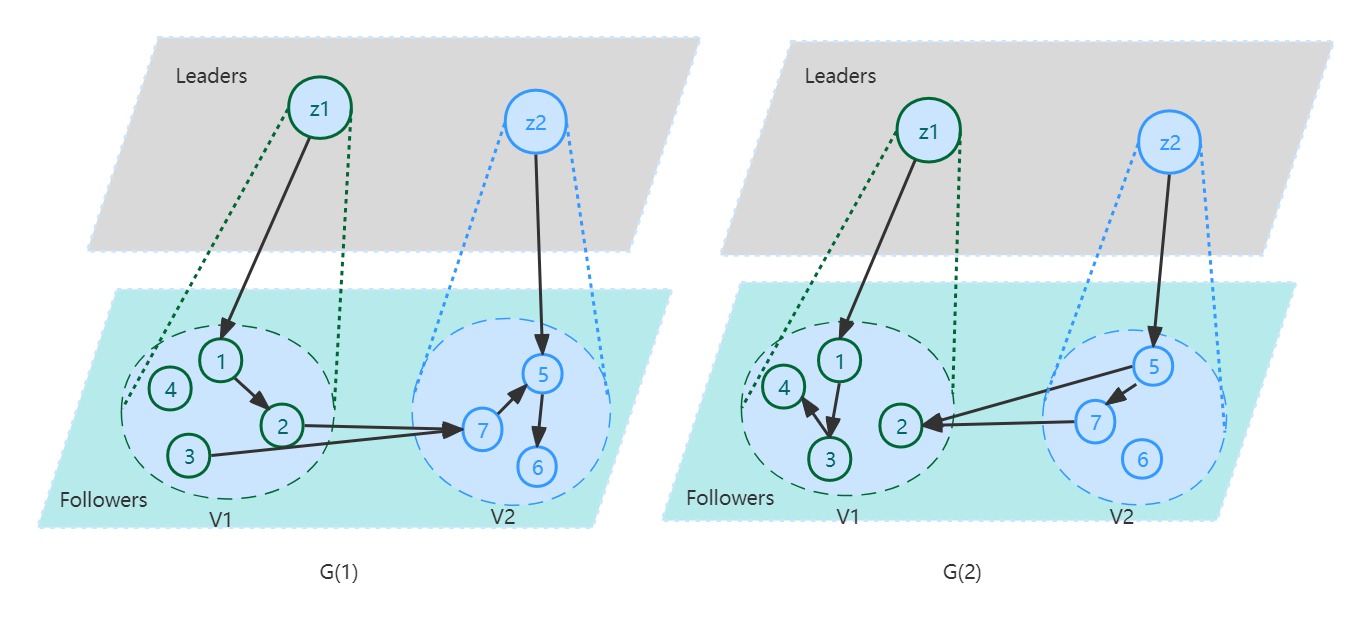}}
\caption{The communication topology contain two interacting clusters.
Each cluster is pinned by a desire trajectory. The union graphs have a spanning tree.}
\label{fig1}
\end{figure}

We separate the systems into two clusters,
$V_1=\{1,2,3,4\}$, and $V_2=\{5,6,7\}$. The leader agents $z_1$ and $z_2$ are also presented
in Fig. \ref{fig1}. The network topology switches every 0.1s.
Let
$$A=
\begin{bmatrix}
  0& 1& 0& 0 \\
  0& 0 &-1& 0 \\
  0& 0 &0 &1\\
  0& 0 &5& 0
\end{bmatrix},
\quad
B=\begin{bmatrix}
    0 \\
    1\\
    0\\
    -2
  \end{bmatrix}
$$

Though calculation, one can find $(A,B)$ is stabiliable;
then, we choose
$$
P=\begin{bmatrix}
 44.1082 &  97.7875 &-137.5021 & -46.9156 \\
   97.7875&  227.4021& -323.4099 &-120.6225  \\
 -137.5021& -323.4099 & 463.3337&  180.8030  \\
  -46.9156& -120.6225 & 180.8030&  101.7587  \\
\end{bmatrix},$$
and $K=\begin{bmatrix}
18.7954&   51.0090 & -79.3369 & -69.8915
\end{bmatrix} $.

We first set up the initial conditions of the agents and the network topology.
We then execute the cluster synchronization algorithm iteratively until convergence.
We measure the convergence time, which is defined as the number of
iterations required for all clusters to reach cluster consensus.
We choose  the initial states of each agent from $[-10,10] \times[-10,10]$.
Then, we introduce $E_1(t)=\sum_{i=1}^4\left\|x_i(t)-s_1(t)\right\|, \quad E_2(t)=$ $\sum_{i=5}^7\left\|x_i(t)-s_2(t)\right\|$  describe the error
between agent $1,2, 3, 4$ and $s_1$, agent $5,6,7$ and $s_2$.

\begin{figure}[!t]
\centerline{\includegraphics[width=9cm]{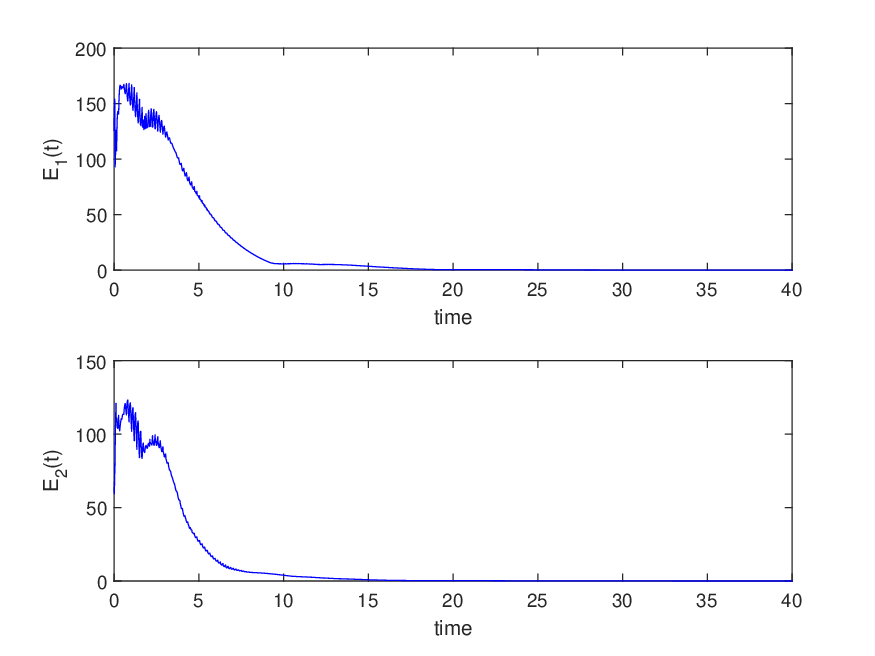}}
\caption{The trajectories of $E_1(t)$ and $E_2(t)$ with $\omega$=0.5, $c_1$=2, $c_2=2$.}
\label{fig_error2}
\end{figure}

\begin{figure}[!t]
\centerline{\includegraphics[width=9cm]{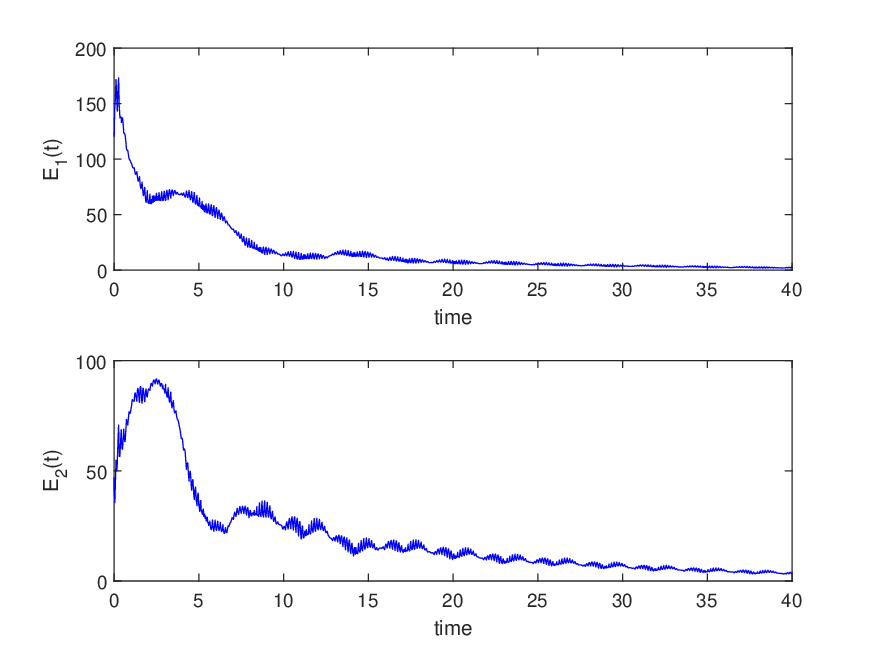}}
\caption{The trajectories of $E_1(t)$ and $E_2(t)$ with $\omega$=0.8, $c_1$=2, $c_2=2$.}
\label{fig_error3}
\end{figure}

\begin{figure}[!t]
\centerline{\includegraphics[width=9cm]{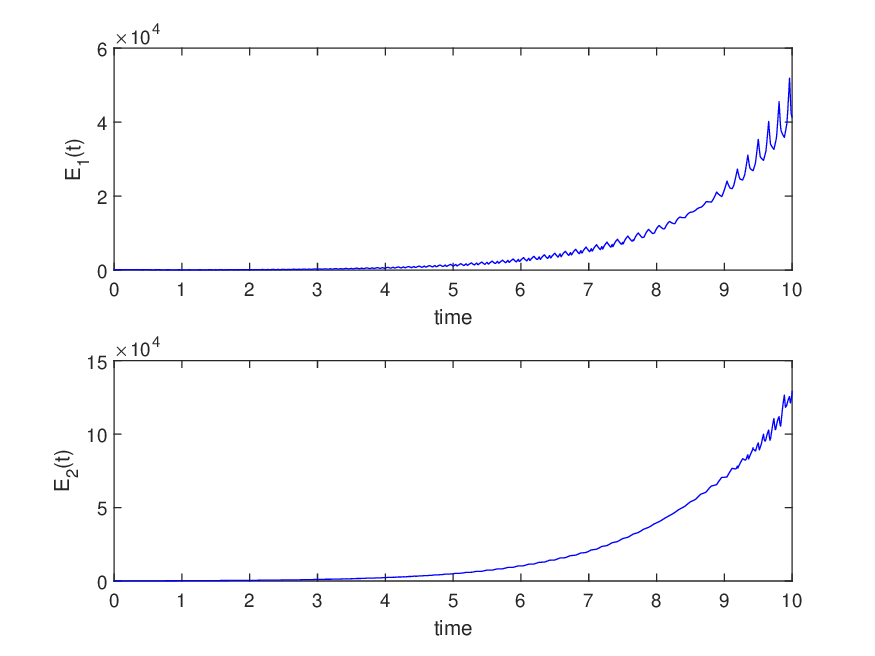}}
\caption{The trajectories of $E_1(t)$ and $E_2(t)$ with $\omega$=1, $c_1$=2, $c_2=2$.}
\label{fig_error4}
\end{figure}

\begin{figure}[!t]
\centerline{\includegraphics[width=9cm]{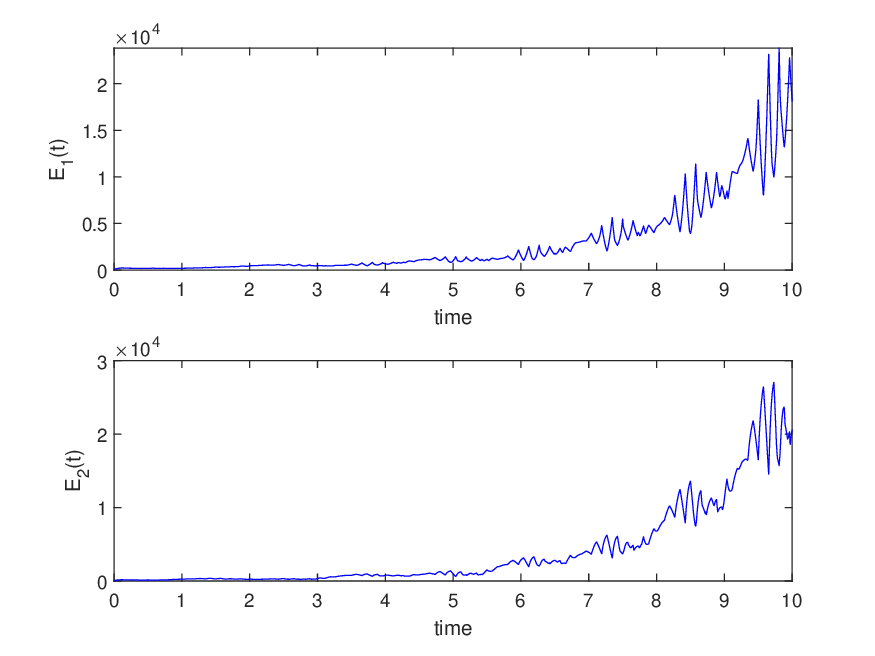}}
\caption{The trajectories of $E_1(t)$ and $E_2(t)$ with $\omega$=0.8, $c_1$=2, $c_2=2$
and state matrices $(A,B)$ unstable.}
\label{fig_error5}
\end{figure}

From Fig. \ref{fig_error2},
one can find if we choose the coupling strength as
$\omega=0.5, c_1=2 \geq-\left(\left[\lambda_{\min }\left(\Xi \bar{L}_0^{\infty}+\right.\right.\right.$
$\left.\left.\left.\bar{L}_0^{\infty T} \Xi\right)\right] /\left[\lambda_{\min }\left(\Xi_1
\bar{L}_{11}^{\infty}+\bar{L}_{11}^{\infty T} \Xi_1\right)\right]\right)=0.8643$, and
$c_2=2 \geq
\frac{\lambda_{\min }\left(\Xi \bar{L}_0^{\infty}+\bar{L}_0^{\infty T} \Xi\right)}{\lambda_{\min}(\Xi_2 \bar{L}_{22}^{\infty}+\bar{L}_{22}^{\infty T}\Xi_{2})}=0.9567,
$
the cluster synchronization can be achieved.
From Fig. \ref{fig_error3}, one can find the coupled linear systems will achieve cluster
synchronization if $\alpha=0.8 , c_1=2$, and $c_2=2$.
Compare with Fig. \ref{fig_error2}, the speed of achieve cluster synchronization is lower.

Then, if we choose $\omega=1$, one can find the coupled system can not achieve cluster
synchronization, see Fig. \ref{fig_error4}
If the topology swithing form Fig.\ref{fig2}, one can find the coupled
system will no achieve consensus, see Fig. \ref{fig_error5}.

\begin{figure}[!t]
\centerline{\includegraphics[width=9cm]{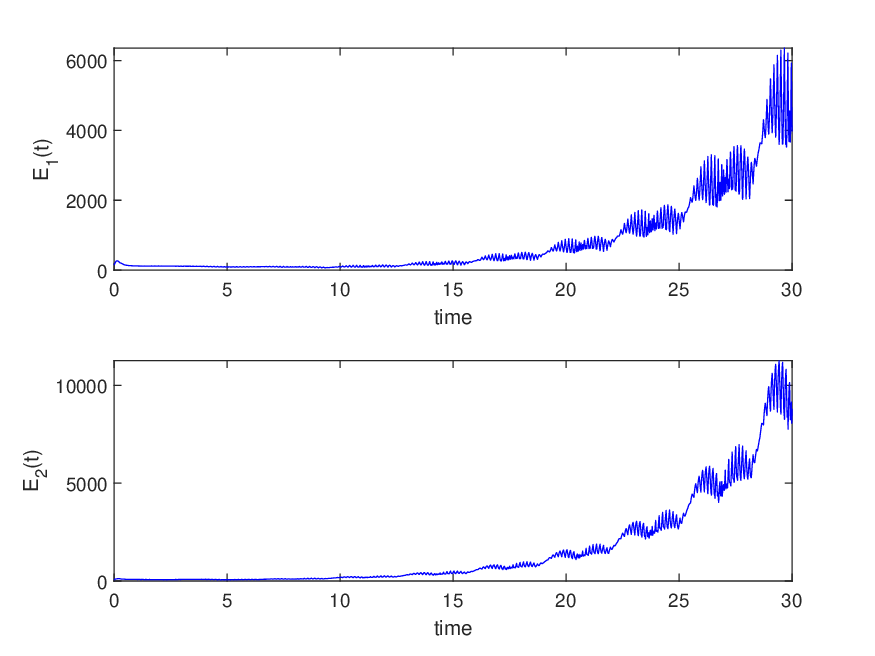}}
\caption{The trajectories of $E_1(t)$ and $E_2(t)$ with $\omega$=0.5, $c_1$=2, $c_2=2$ and union graph does not has a spanning tree.}
\label{fig_error6}
\end{figure}

If we choose  matrices as

$$A^{'}=
\begin{bmatrix}
  0& 1& 0& 0 \\
  0& 0 &-1& 0 \\
  0& 0 &0 &1\\
  0& 0 &0& 5
\end{bmatrix},
\quad
B^{'}=\begin{bmatrix}
    0 \\
    1\\
    0\\
    -2
  \end{bmatrix}
$$

By calculation, the matrices $(A^{'},B^{'})$ is not contrllable, one can find  coupled system
can not achieve cluster synchronization, see Fig. \ref{fig_error6}.

Our simulation results demonstrate that the proposed cluster synchronization algorithm performs
effectively over switching network topologies. The algorithm
achieves fast convergence and high synchronization accuracy.

\begin{figure}[!t]
\centerline{\includegraphics[width=9cm]{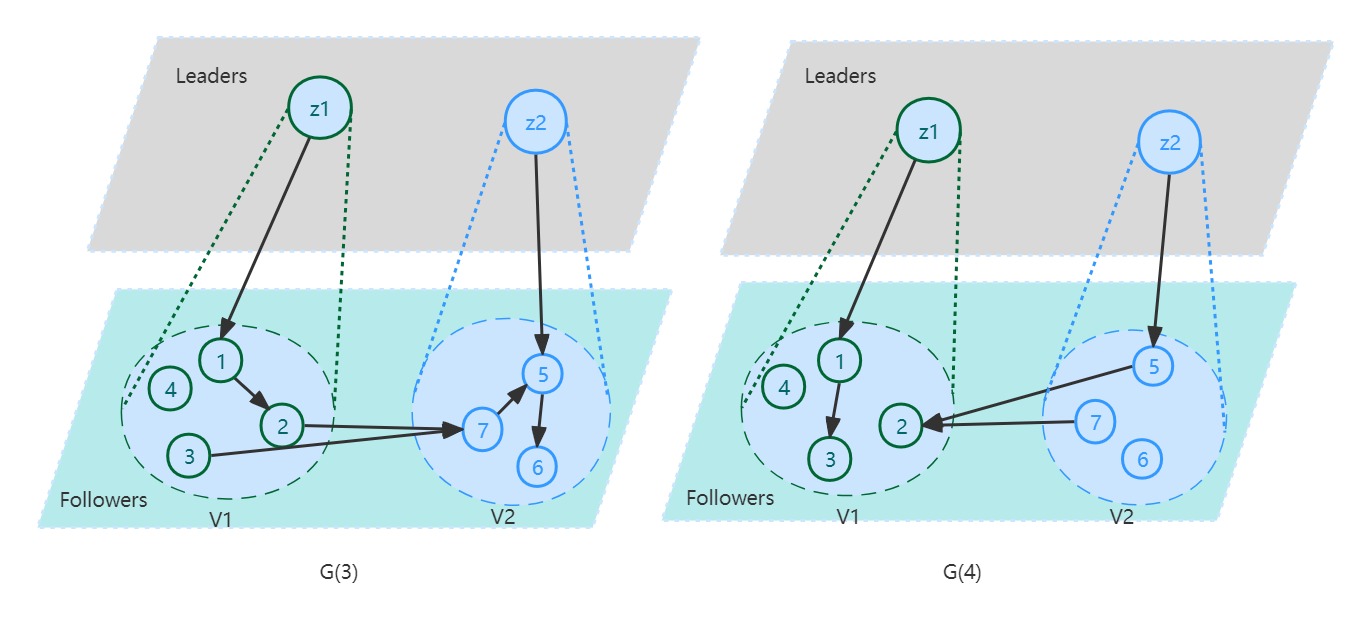}}
\caption{The communication topology contain two interacting clusters.
Each cluster is pinned by a desire trajectory. The union graphs do not have a spanning tree.}
\label{fig2}
\end{figure}

\end{example}


\section{Conclusion}\label{sec_con}
In summary, this paper has delved into the intricate realm of Cluster Synchronization amidst the challenges posed by a rapidly switching network topology within the zero trust framework.
The preliminary section established the grounding the study in Graph Theory and a comprehensive system model.
Moving forward, the core of our investigation in Section III addressed Cluster Synchronization within a linear system. We identified both necessary and sufficient conditions for achieving consensus among multiple agents in the dynamic landscape of fast-switching topologies. These conditions provide a robust theoretical foundation for understanding and implementing synchronization strategies.
The subsequent, rigorously presented the proofs of the main results, validating the theoretical assertions made earlier. This step is crucial in establishing the reliability and applicability of our proposed conditions for achieving cluster consensus.

Furthermore, the simulation results in Section V not only validated the theoretical findings but also demonstrated the practical feasibility and effectiveness of the proposed approach. By bridging the gap between theory and application, our simulations provide valuable insights into the real-world implications of our cluster synchronization framework.

In conclusion, this research contributes to the evolving landscape of secure multi-agent systems by offering a comprehensive approach to cluster synchronization under fast-switching network topologies within the
zero trust framework. As we navigate an era of dynamic technological changes, the insights provided in this paper can serve as a roadmap for the development of robust and secure systems, fostering trust even in environments characterized by constant network fluctuations.

\section*{Reference}

\bibliographystyle{elsarticle-num}
\bibliography{mybibfile}

%
%

\end{document}